\documentclass[prd,showpacs,preprintnumbers,amsmath,amssymb,superscriptaddress,floatfix,nofootinbib]{revtex4}
\usepackage{graphicx}
\usepackage{epstopdf}
\usepackage{amsmath}
\usepackage{amsfonts}
\usepackage{amssymb}
\usepackage{bm}

\begin{document}

\title{ Combined theoretical study of the $D^+ \to \pi^+ \eta \eta$  and $D^+ \to  \pi^+ \pi^0 \eta $ reactions}

\author{Natsumi Ikeno}
\email{ikeno@tottori-u.ac.jp}
\affiliation{Department of Agricultural, Life and Environmental Sciences, Tottori University, Tottori 680-8551, Japan}
\affiliation{Departamento de F\'{\i}sica Te\'orica and IFIC,
Centro Mixto Universidad de Valencia-CSIC Institutos de Investigaci\'on de Paterna, Aptdo.22085, 46071 Valencia, Spain}

\author{Melahat Bayar}
\email{melahat.bayar@kocaeli.edu.tr}
\affiliation{Department of Physics, Kocaeli University, Izmit 41380, Turkey}

\author{Eulogio Oset}
\email{oset@ific.uv.es}
\affiliation{Departamento de F\'{\i}sica Te\'orica and IFIC,
Centro Mixto Universidad de Valencia-CSIC Institutos de Investigaci\'on de Paterna, Aptdo.22085, 46071 Valencia, Spain}
\date{\today}

\begin{abstract}
We study the $D^+ \to \pi^+ \eta \eta $ and $D^+ \to \pi^+ \pi^0 \eta$ reactions, which are single Cabibbo suppressed and can proceed both through internal and external emission. The primary mechanisms at quark level are considered, followed by hadronization to produce three mesons in the $D^+$ decay, and after that the final state interaction of these mesons leads to the production of the $a_0(980)$ resonance, seen in the $\pi^+ \eta$, $\pi^0 \eta$ mass distributions. The theory has three unknown parameters to determine the shape of the distributions and the ratio between the $D^+ \to \pi^+ \eta \eta$ and $D^+ \to \pi^+ \pi^0 \eta$ rates. This ratio restricts much the sets of parameters but there is still much freedom leading to different shapes in the mass distributions. We call for a measurement of these mass distributions that will settle the reaction mechanism, while at the same time provide relevant information on the way that the $a_0(980)$ resonance is produced in the reactions.
\end{abstract}


\maketitle
\section{Introduction}
The $D$ weak decays into three light mesons have proved to be very rich, allowing one to dig into the weak reaction mechanism~\cite{ellis,matsuda,nakagawa} as well as providing information on the meson-meson interaction~\cite{aitala,focus,klempt,babar,babar2,rosner,muramatsu,kappa,ollerkappa,patricia,kubis,robert,dosreis,lroca}.
A review on this latter issue can be seen in Ref.~\cite{review}. The rich field of meson-meson interactions gives rise to many mesonic resonances that show up in most of the $D$ decays. Normally, the possible pairs with the tree final mesons lead to some mesonic resonances and it is common to see the effects of several resonances in just one decay. For instance, in the $D^0 \to K^- \pi^+ \eta$ reaction~\cite{belle}, one has the contributions of the $a(980)$, $\kappa (K^*_0(700))$ and $K^*(890)$, among other resonances that only have a minor effect in the mass distributions~\cite{toledoikeno}.
Another example would be the $D^0 \to \eta \pi^+ \pi^-$ reaction measured at BESIII~\cite{besdecay}, where one expects contribution from the $\rho^0$, $a_0(980)$, $f_0(500)$ and $f_0(980)$ resonances. It is interesting to look for reactions which show only effects of a single resonance. This was the case of the $D^+_s \to \pi^+ \pi^0 \eta$ reaction measured at BESIII~\cite{besds}, which was shown to be dominated by the $a_0(980)$, seen in the $\pi^+ \eta$ and $\pi^0 \eta$ mass distribution~\cite{raquel}.
Actually in this reaction, the $\pi^+ \pi^0$ can also come from the $\rho^+$, but in the experiment~\cite{besds} this contribution was eliminated with a simple cut, demanding that $M(\pi^+ \pi^0) > 1~$GeV.

In the present paper, we want to study two reactions, also measured at Ref.~\cite{besdecay} although without information on mass distributions: the $D^+ \to \pi^+ \pi^0 \eta$ and $D^+ \to \pi^+ \eta \eta$ reactions. In the first one we expect to have contributions of the $\rho^+$ and $a_0(980)$ and in the second one of the $a(980)$. As in Ref.~\cite{besds}, the $\rho^+$ contribution in the $D^+ \to \pi^+ \pi^0 \eta$ reaction can be eliminated with the same cut, requiring that $M(\pi^+ \pi^0)>1$~GeV, and then the two reactions can be related and will show only the $a_0(980)$ resonance. 
Actually, one of these reactions, the $D^+ \to \pi^+ \pi^0 \eta$, has been studied theoretically in Ref.~\cite{enwang} and shown to be Cabibbo suppressed, exhibiting clearly the $a_0(980)$ excitation. We plan to study the two reactions together, along the lines of Ref.~\cite{enwang}, using the extra information provided in Ref.~\cite{besdecay} about the ratio of branching fractions of these two decay modes. This information puts constraints on the free parameters of the theory and allows us to make predictions on mass distributions which can be tested in future runs of the reactions with more statistics.

\section{Formalism}\label{sec:form}
Although the final state in the $D^+ \to \pi^+ \pi^0 \eta$ reaction is the same as in $D^+_s \to \pi^+ \pi^0 \eta$  measured in Ref.~\cite{besds}, the reaction mechanism is different here since it allows contributions from external emission and internal emission~\cite{chau}, while in the $D^+ _s \to \pi^+ \pi^0 \eta$ external emission did not occur and the process proceeded via internal emission~\cite{raquel}. 
In addition, the $D^+ \to \pi^+ \pi^0 \eta$ reaction is Cabibbo suppressed and there are two topologically different mechanisms to proceed for each of the different modes. This necessarily introduces more unknowns in the theoretical description, in spite of which it was shown in Ref.~\cite{enwang} that the $a_0(980)$ signal should show up clearly in the reaction. The data on the branching ratios for $D^+ \to \pi^+ \pi^0 \eta$ and $D^+ \to \pi^+ \eta \eta$ will limit somewhat the theoretical freedom and allow us to make more constraint predictions in both reactions. 

\begin{figure}[htb!]
  \centering
  \includegraphics[width=0.7\textwidth]{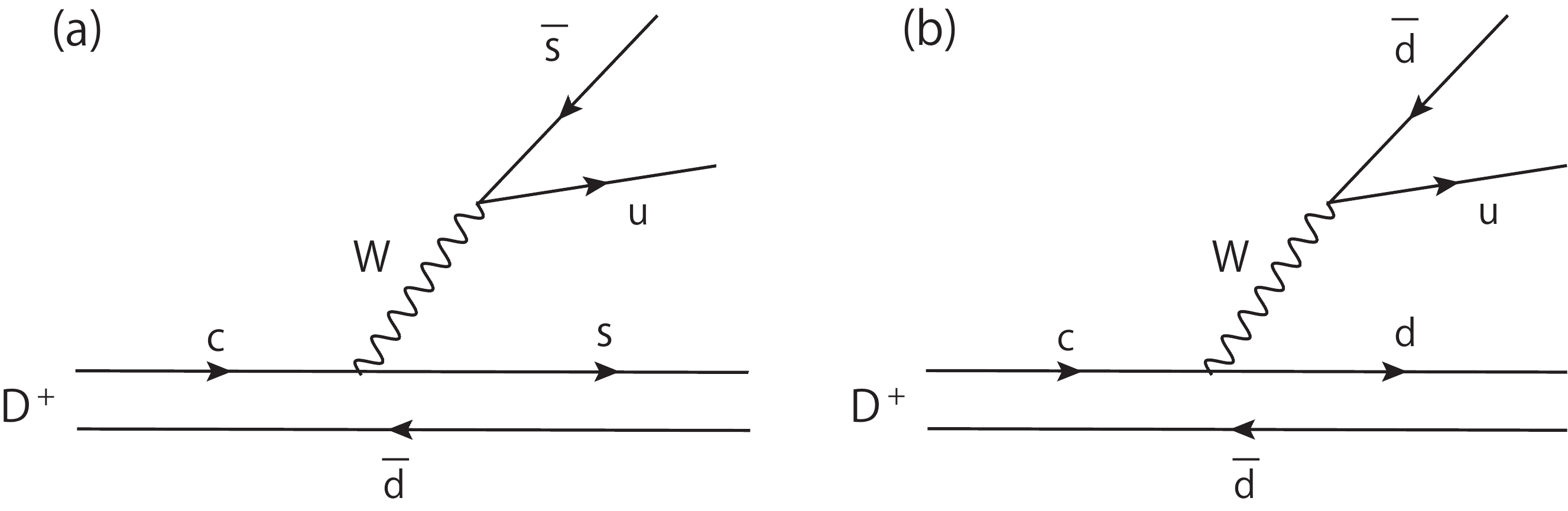}
  \caption{ Diagrams of external emission at the quark level: (a) Cabibbo suppressed $Wu \bar s$ vertex, (b) Cabibbo suppressed $Wcd$ vertex.
 }
  \label{fig:1}
\end{figure}

Let us begin with external emission. We can have the primary mechanisms at the quark level depicted in Fig.~\ref{fig:1}.

\begin{figure}[htb!]
  \centering
  \includegraphics[width=0.7\textwidth]{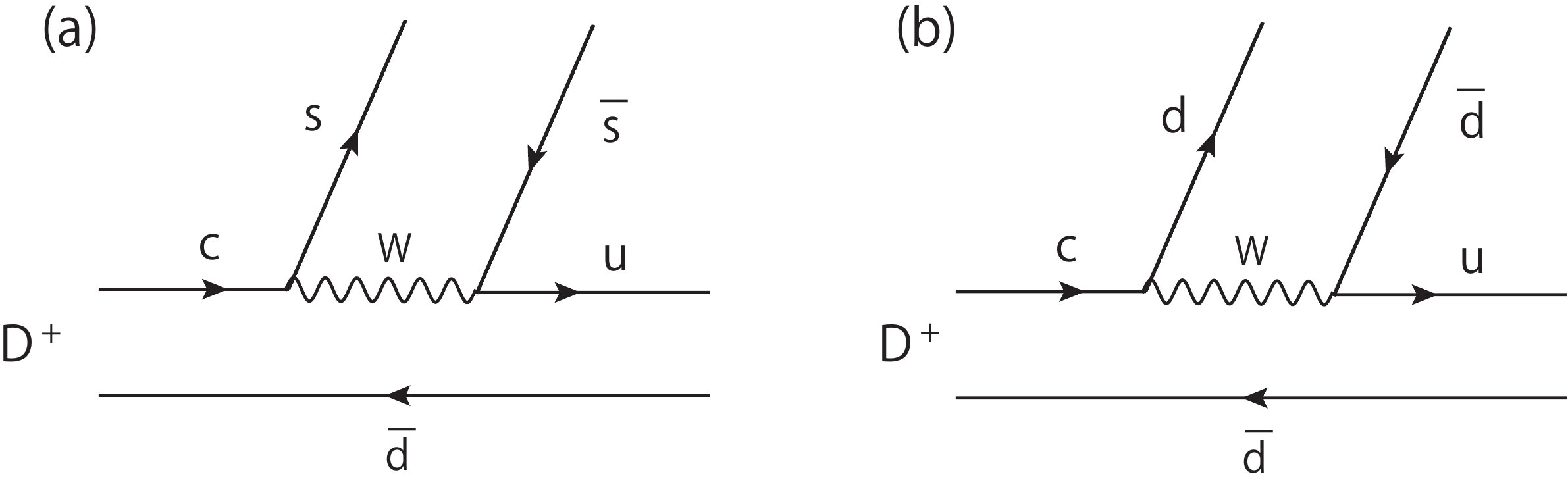}
  \caption{ Diagrams for internal emission at the quark level: (a) Cabibbo suppressed $W \bar s u$ vertex, (b) Cabibbo suppressed $Wcd$ vertex. }
  \label{fig:2}
\end{figure}

At the same time we can also have internal emission which is depicted in Fig.~\ref{fig:2}.

\begin{figure}[htb!]
 \centering
 \includegraphics[width=0.7\textwidth]{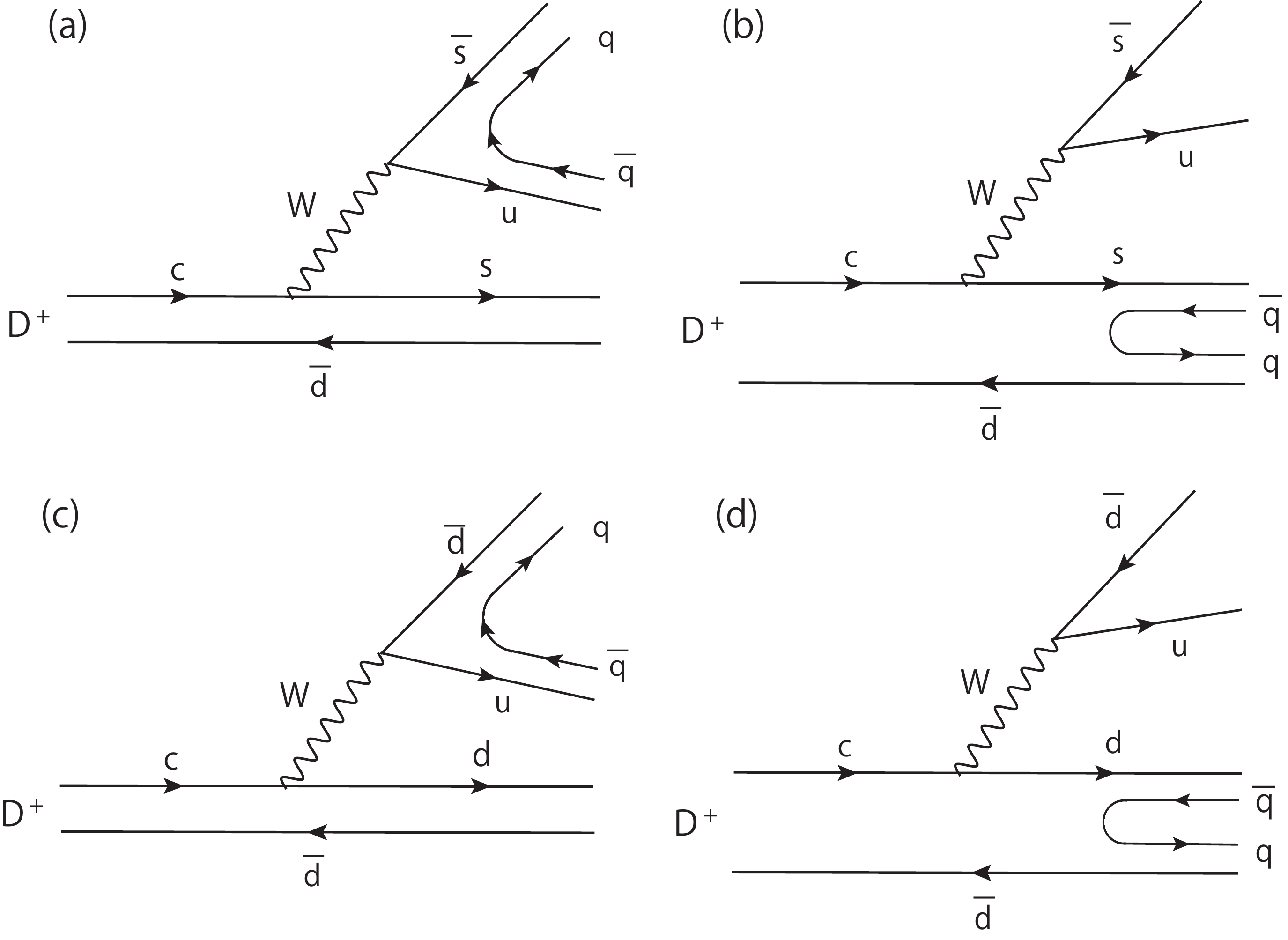}
 \caption{ Hadronization in the diagrams of Fig.~\ref{fig:1}. }
 \label{fig:3}
\end{figure}

\begin{figure}[htb!]
 \centering
 \includegraphics[width=0.7\textwidth]{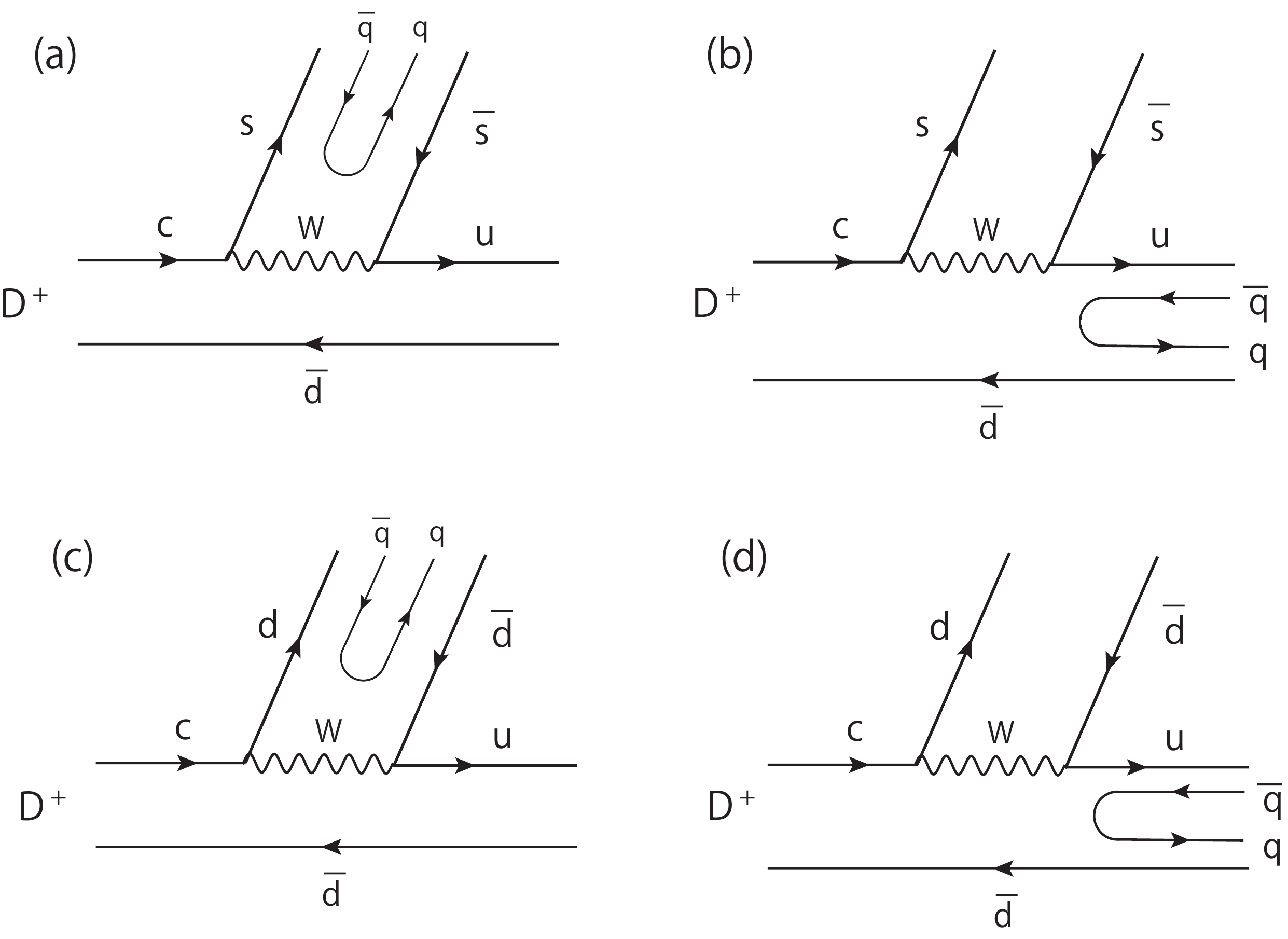}
 \caption{ Hadronization in the diagrams of Fig.~\ref{fig:2}.  }
\label{fig:4}
\end{figure}

Next we proceed to hadronize those mechanisms introducing a $\bar q q$ pair SU(3) singlet $\bar u u + \bar d d + \bar s s $, which is depicted in Figs.~\ref{fig:3} and \ref{fig:4}.

In Fig.~\ref{fig:3}(a), we will have the hadronization of the $u \bar s$ pair as
\begin{equation}
 u \bar s  \to  \sum_i u \bar{q}_i q_i \bar{s} = \sum_i M_{1i} M_{i3 } = ( M^2 )_{13},
\end{equation}   
 where $M$ is the $q\bar{q}$ matrix, which we write in terms of physical mesons as  
\begin{equation}
  M \to P \equiv 
\left(  
\begin{array}{ccc}
\frac{\pi^0}{\sqrt{2}}+\frac{\eta}{\sqrt{3}}
 & \pi^+ & K^+ \\
\pi^- & -\frac{\pi^0}{\sqrt{2}}+\frac{\eta}{\sqrt{3}}  & K^0  \\
K^- & \bar{K}^0 & -\frac{\eta}{\sqrt{3}}
\end{array}    
\right)\,
\label{pmatrix}
 \end{equation} 
where we have used the standard $\eta$, $\eta^\prime$ mixing of Ref.~\cite{bramon} and omitted the $\eta'$ which has a very high mass and does not play a role in the generation of the low energy scalar resonances~\cite{npa}.

The mesons that appear in the hadronization are
\begin{eqnarray} 
& & {\rm Fig}.~\ref{fig:3}(a):~~  ( M^2 )_{13} \bar K^0 = \left( \frac{\pi^0 K^+}{\sqrt{2}} + \pi^+ K^0 \right) \bar K^0, \\
& & {\rm Fig}.~\ref{fig:3}(b):~~  ( M^2 )_{32} K^+ = \left(  K^- \pi^+ -\frac{\pi^0 \bar K^0}{\sqrt{2}}  \right) K^+, \\
& & {\rm Fig}.~\ref{fig:3}(c):~~  ( M^2 )_{12} \left( -\frac{\pi^0}{\sqrt{2}}+\frac{\eta}{\sqrt{3}} \right) 
= \left(  \frac{2}{\sqrt{3}}\eta \pi^+ + K^+ \bar K^0  \right) \left( -\frac{\pi^0}{\sqrt{2}}+\frac{\eta}{\sqrt{3}} \right), \\
& & {\rm Fig}.~\ref{fig:3}(d):~~  ( M^2 )_{22} \pi^+ = \left(  \pi^- \pi^+  + \frac{\pi^0 \pi^0}{2} + \frac{\eta \eta}{3}  -\frac{2}{\sqrt{6}}\pi^0 \eta + K^0 \bar K^0  \right) \pi^+, \\
& &{\rm Fig}.~\ref{fig:4}(a):~~  ( M^2 )_{33} \pi^+ = \left(  K^- K^+  + \bar K^0  K^0  + \frac{\eta \eta}{3}  \right) \pi^+, \\
& & {\rm Fig}.~\ref{fig:4}(b):~~  ( M^2 )_{12} \left( -\frac{\eta}{\sqrt{3}} \right) = \left(  \frac{2}{\sqrt{3}}\eta \pi^+ + K^+ \bar K^0  \right) 
 \left( -\frac{\eta}{\sqrt{3}}  \right) ,\\
& &{\rm Fig}.~\ref{fig:4}(c):~~  ( M^2 )_{22} \pi^+ = \left(  \pi^- \pi^+  + \frac{\pi^0 \pi^0}{2} + \frac{\eta \eta}{3}  -\frac{2}{\sqrt{6}}\pi^0 \eta + K^0 \bar K^0  \right) \pi^+, \\
& & {\rm Fig}.~\ref{fig:4}(d):~~  ( M^2 )_{12} \left( -\frac{\pi^0}{\sqrt{2}}+\frac{\eta}{\sqrt{3}} \right) 
= \left(  \frac{2}{\sqrt{3}}\eta \pi^+ + K^+ \bar K^0  \right) \left( -\frac{\pi^0}{\sqrt{2}}+\frac{\eta}{\sqrt{3}} \right) .
\end{eqnarray}

Since Figs.~\ref{fig:3}(a),~\ref{fig:3}(c) have the same topology and the same Cabibbo suppressing factor, we can sum them and the same can be said about Figs.~\ref{fig:3}(b),~\ref{fig:3}(d). For internal emission, we can also sum the contributions of Figs.~\ref{fig:4}(a),~\ref{fig:4}(c) and \ref{fig:4}(b),~\ref{fig:4}(d).

However, there is a subtlety about the diagrams of type Figs.~\ref{fig:3}(a),~\ref{fig:3}(c). The effective $WPP$ ($P$ pseudoscalar meson) vertex can be evaluated with effective chiral Lagrangians  $W^\mu \langle [P, \partial_\mu P] T_{-} \rangle$ with $W^\mu$ the $W$ field and $ T_{-}$ a matrix related to the Cabibbo-Kobayashi-Maskawa elements~\cite{gasser,scherer}.
If we wish to get the two pseudoscalar mesons in $s$-wave, which we need to produce the scalar resonances, we get such a contribution with this Lagrangian with $\mu =0$, which produces a vertex proportional to $p^0 - p'^0$ in the rest frame of $W$, and hence vanishes for particles with equal mass. This is discussed in Ref.~\cite{pedrosun}. This means we can keep the $\pi K$ and $\pi \eta$ terms in Fig.~\ref{fig:3}(a), \ref{fig:3}(c) but we must omit the $K^+ \bar K^0$ contribution in Fig.~\ref{fig:3}(c). These mechanisms were neglected in Ref.~\cite{enwang}, but we keep them here and play with the relative weight with respect to the hadronization in Fig.~\ref{fig:3}(b), \ref{fig:3}(d) in order to get agreement with the branching ratios of $D^+ \to \pi^+ \pi^0 \eta$ and $D^+ \to \pi^+ \eta \eta$~\cite{besdecay}.

This said, we obtain from hadronization of the upper $q \bar q$ pair of external emission in Figs.~\ref{fig:3}(a), \ref{fig:3}(c) the hadronic contribution
\begin{equation}
  H_1 = \pi^+ K^0 \bar K^0 - \frac{2}{\sqrt{6}} \eta \pi^+ \pi^0 + \frac{2}{3} \eta \eta \pi^+ + \frac{1}{ \sqrt{2}} \pi^0 K^+ \bar K^0,
\end{equation}
and from hadronization of the lower $q \bar q$ pair (Figs.~\ref{fig:3}(b), \ref{fig:3}(d))
\begin{equation}
H_2 = K^+ K^- \pi^+ + K^0 \bar K^0 \pi^+ - \frac{1}{\sqrt{2}} \pi^0 \bar K^0 K^+ + \pi^+ \pi^- \pi^+ +\frac{1}{2} \pi^0 \pi^0 \pi^+ + \frac{1}{3} \eta \eta \pi^+ - \frac{2}{\sqrt{6}} \pi^0 \eta\pi^+.
\end{equation}
Similarly, summing the contributions of Figs.~\ref{fig:4}(a), \ref{fig:4}(c) for internal emission, we find
\begin{equation}
 H'_1 = K^- K^+ \pi^+ + 2 K^0 \bar K^0 \pi^+ + \frac{2}{3} \eta \eta \pi^+ + \pi^- \pi^+ \pi^+ + \frac{1}{2} \pi^0 \pi^0 \pi^+ - \frac{2}{\sqrt{6}}\pi^0 \eta \pi^+,
\end{equation}
and summing the contribution of Figs.~\ref{fig:4}(b), \ref{fig:4}(d)
\begin{equation}
 H'_2 = -\frac{2}{\sqrt{6}} \eta \pi^0 \pi^+ - \frac{1}{\sqrt{2}} \pi^0 K^+ \bar K^0.
\end{equation}

We can further simplify these expressions. Since $K^- K^+ + K^0 \bar K^0$ has isospin $I=0$, the combination $ (K^- K^+ + K^0 \bar K^0) \pi^+$ cannot give rise to $\pi^0 \eta \pi^+$ upon rescattering since $\pi^0 \eta$ has $I=1$. It could contribute to $\eta \eta \pi^+$ production but $\eta \eta$ is far away from the narrow $f_0(980)$ resonance and will also be ineffective in this channel. Similar arguments can be done with the contribution $\pi^+ \pi^- + \frac{\pi^0 \pi^0}{2}$ which has $I=0$ and hence cannot give $\pi^0 \eta$ upon rescattering. It could give $\eta \eta$ but the same argument as above holds. This means that far practical purposes we can write
\begin{equation}
H_2 \equiv - \frac{1}{\sqrt{2}} \pi^0 \bar K^0 K^+ + \frac{1}{3} \eta \eta \pi^- - \frac{2}{\sqrt{6}} \pi^0 \eta \pi^+.
\end{equation}

The same arguments can be used concerning $H'_1$ and $H'_2$ and we can then rewrite for effective purposes
\begin{equation}
 H'_1 \equiv - \frac{2}{\sqrt{6}} \pi^0 \eta \pi^+ + K^0 \bar K^0 \pi^+ + \frac{2}{3}\eta \eta \pi^+,
\end{equation}
and $H_1$, $H'_2$ are not changed.
We should note that eliminating the $\eta \eta \pi^+$ terms in $H_2$, $H'_1$, $H'_2$, which do not contribute to the $\pi^+ \pi^0 \eta$ production studied in Ref.~\cite{enwang}, we obtain the same terms as in Ref.~\cite{enwang}. In order to keep a close analogy to the results of Ref.~\cite{enwang}, we give weights to the different terms;
\begin{equation}
H_1 \to A \beta,~~~~~ H_2 \to A,~~~~~ H'_1 \to B, ~~~~ H'_2 \to B \gamma.
\end{equation}

As discussed in Ref.~\cite{enwang}, one expects $B \sim \frac{1}{3}A$, as it corresponds to the internal emission color suppression, and $\gamma \sim 1$. We shall also assume values around these ratios taking $A=1 $ with arbitrary normalization.

\begin{figure}[tb!]
  \centering
  \includegraphics[width=0.75\textwidth]{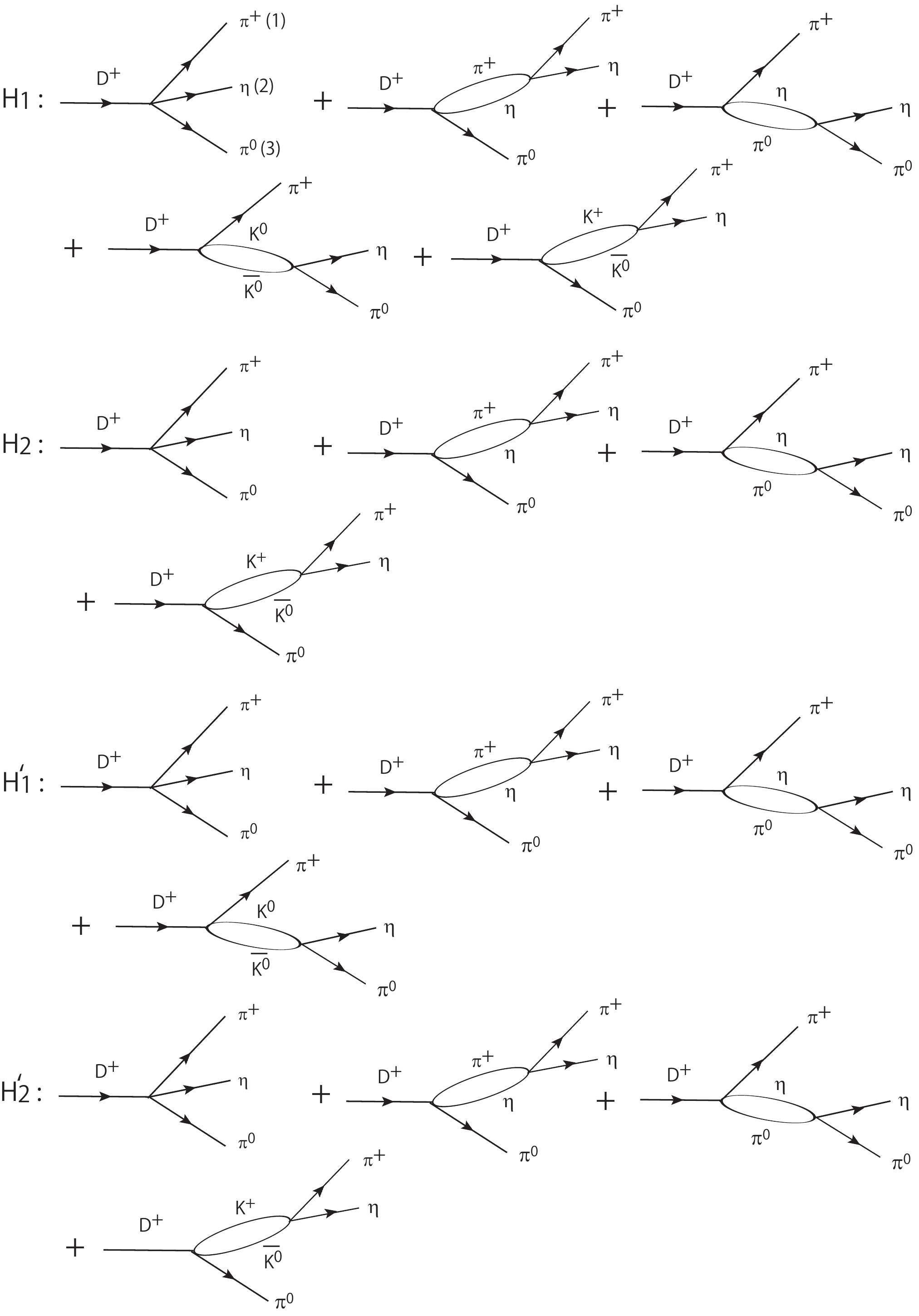}
  \caption{ Rescattering of terms to produce $\pi^+ \pi^0 \eta$.}
  \label{fig:5}
\end{figure}

The next step consists of allowing the interaction of pairs of particles to finally have either the $\pi^+ \pi^0 \eta$ or $\pi^+ \eta \eta$ final state. The possible ways to have $\pi^+ \pi^0 \eta$ are given in Fig.~\ref{fig:5}.

\begin{figure}[tb!]
  \centering
  \includegraphics[width=0.75\textwidth]{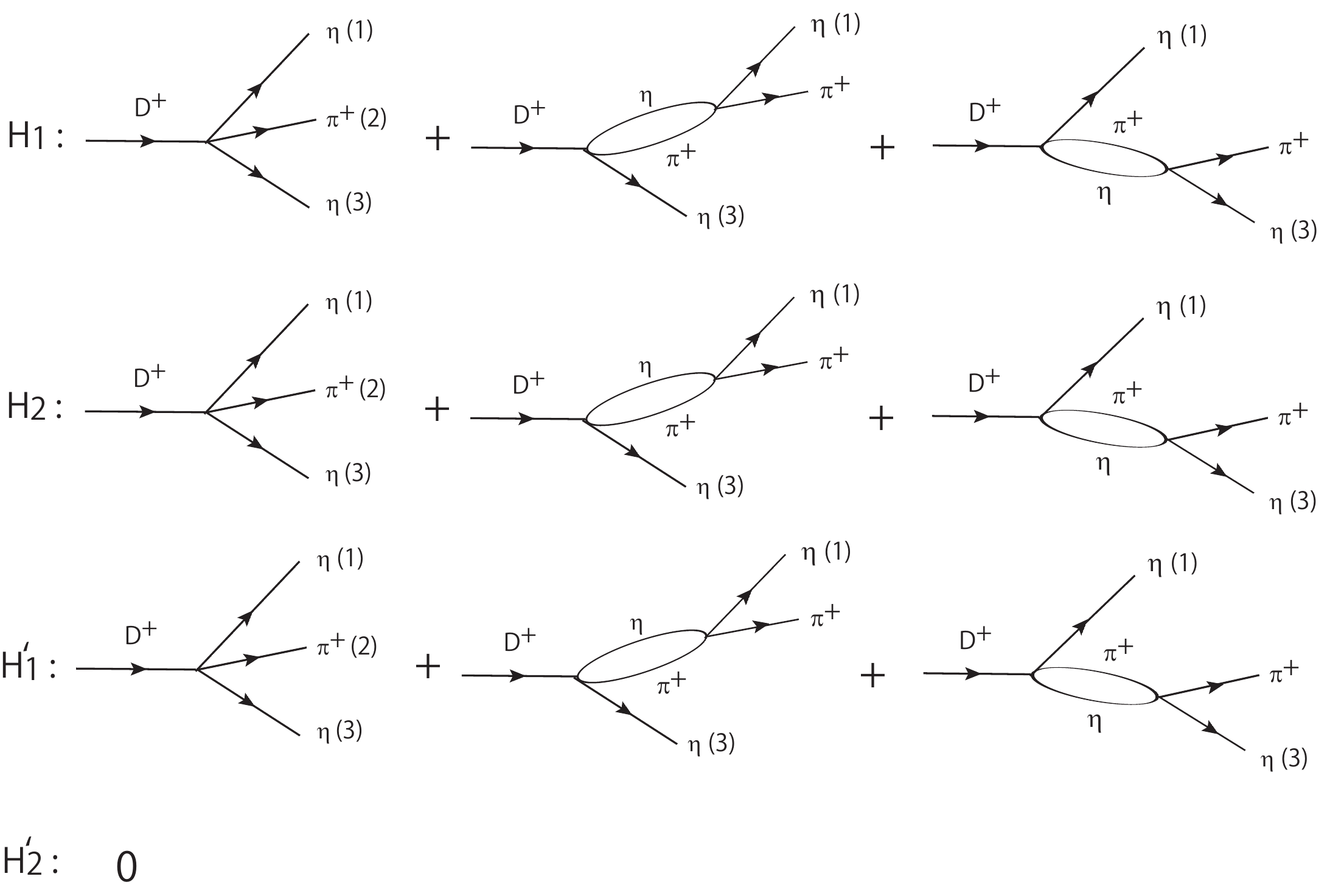}
  \caption{ Rescattering of terms to produce $\pi^+ \eta \eta$.}
  \label{fig:6}
\end{figure}

Similarly, if we wish to produce $\pi^+ \eta \eta$, we will have the mechanisms depicted in Fig.~\ref{fig:6}.

Note that in Fig.~\ref{fig:6} we are neglecting $K^0 \bar K^0 \to \eta \eta$ because the $f_0(980)$ resonance strength already becomes negligible above the $\eta \eta$ threshold. By looking at the diagrams in Figs.~\ref{fig:5},~\ref{fig:6}, we can write the final amplitude for the two reactions. For this purpose we define the weight of the different terms as:
\begin{eqnarray}
& & H_1:~~~ h_{\pi^+ \pi^0 \eta} = -\frac{2}{\sqrt{6}};~~~ h_{\pi^+ K^0 \bar K^0 } = 1;~~~ ~~~ h_{\pi^0 K^+ \bar K^0 } = \frac{1}{\sqrt{2}};~~~ h_{\pi^+ \eta \eta} =\frac{2}{3}, \\
& &  H_2:~~~ \bar h_{\pi^+ \pi^0 \eta} = -\frac{2}{\sqrt{6}};~~~ \bar h_{\pi^0 K^+ \bar K^0 } = -\frac{1}{\sqrt{2}};~~~ \bar h_{\pi^+ \eta \eta} =\frac{1}{3}, \\
& &  H'_1:~~~ h'_{\pi^+ \pi^0 \eta} = -\frac{2}{\sqrt{6}};~~~ h'_{\pi^+ K^0 \bar K^0 } = 1;~~~ h'_{\pi^+ \eta \eta} =\frac{2}{3}, \\
& &  H'_2:~~~ \bar h'_{\pi^+ \pi^0 \eta} = -\frac{2}{\sqrt{6}};~~~ \bar h'_{\pi^0 K^+ \bar K^0 } = -\frac{1}{\sqrt{2}}.
\end{eqnarray}

And the amplitudes are now written as,
\begin{eqnarray}
t_{D^+ \to \pi^+ \pi^0 \eta} &=&  
\left(  
 h_{\pi^+ \pi^0 \eta} A \beta + \bar h_{\pi^+ \pi^0 \eta} A + h'_{\pi^+ \pi^0 \eta} B + \bar h'_{\pi^+ \pi^0 \eta} B\gamma
\right) \nonumber\\
& & \cdot \left( 
1 + G_{\pi \eta}(M_{\rm inv}(\pi^+ \eta))\ t_{\pi^+ \eta, \pi^+ \eta}(M_{\rm inv}(\pi^+ \eta))
+ G_{\pi \eta}(M_{\rm inv}(\pi^0 \eta))\ t_{\pi^0 \eta, \pi^0 \eta}(M_{\rm inv}(\pi^0 \eta))
\right) \nonumber\\
&+ & 
\left(  
 h_{\pi^+ K^0 \bar K^0} A \beta +  h'_{\pi^+ K^0 \bar K^0} B \right) 
 G_{K \bar K}(M_{\rm inv}(\pi^0 \eta))\ t_{K^0 \bar K^0, \pi^0 \eta}(M_{\rm inv}(\pi^0 \eta))
\nonumber\\
&+ & 
\left(  
 h_{\pi^0 K^+ \bar K^0} A \beta +  \bar h_{\pi^0  K^+ \bar K^0} A +  \bar h'_{\pi^0  K^+ \bar K^0} B \gamma \right) 
  G_{K \bar K}(M_{\rm inv}(\pi^+ \eta))\ t_{K^+ \bar K^0, \pi^+ \eta}(M_{\rm inv}(\pi^+ \eta)), 
\label{eq:T_pipieta}
\\
t_{D^+ \to \pi^+ \eta \eta} &=&  
\frac{2}{\sqrt{2}}
\left(  
 h_{\pi^+ \eta \eta} A \beta + \bar h_{\pi^+ \eta \eta} A + h'_{\pi^+ \eta \eta} B
\right) \nonumber\\
& & \cdot \left( 
1 + G_{\pi \eta}(M_{\rm inv}(\pi^+ \eta(1)))\ t_{\pi^+ \eta, \pi^+ \eta}(M_{\rm inv}(\pi^+ \eta(1))) + G_{\pi \eta}(M_{\rm inv}(\pi^+ \eta(3)))\ t_{\pi^+ \eta, \pi^+ \eta}(M_{\rm inv}(\pi^+ \eta(3)))
\right),
\label{eq:T_pietaeta}
\nonumber\\
\end{eqnarray}
where in $t_{D^+ \to \pi^+ \eta \eta}$ we have taken into account the factor $2$ of symmetry in the amplitude because of the two $\eta$ and included a factor $\frac{1}{\sqrt{2}}$ such that when squaring $t$ we get the factor $\frac{1}{2}$ of symmetry in the width for two identical particles in the final state.

The differential width, up to an arbitrary normalization, is given by Ref.~\cite{pdg},
\begin{equation}
\frac{d^2 \Gamma}{dM_{\rm inv}(12) dM_{\rm inv}(23)}=\frac{1}{(2\pi)^3}
\frac{M_{\rm inv}(12) M_{\rm inv}(23)}{8 m_{D^{+}}^{3}} \ |t|^2 ,
\label{dGam}
\end{equation}    
and according to Figs.~\ref{fig:5},~\ref{fig:6} we have $\pi^+(1)$, $\eta(2)$, $\pi^0(3)$ for $\pi^+ \pi^0 \eta$ production and $\eta(1)$, $\pi^+(2)$, $\eta(3)$ for $\pi^+ \eta \eta$ production. 
 The single differential mass distribution $\displaystyle \frac{d \Gamma}{dM_{\rm inv}(12)}$ is obtained integrating over $M_{23}$ with the limits of the Dalitz plot that are shown in the PDG~\cite{pdg}. 

The scattering matrices are calculated using the chiral unitary approach~\cite{npa} with the coupled channels, $K^+ K^-$, $K^0 \bar K^0$, $\pi^0 \eta$ using
\begin{equation}
T= [ 1- V G]^{-1} V,
\end{equation}
with the potential $V_{ij}$ between the channels given in Ref.~\cite{daixie} and the $G$ function regularized with a cut off $q_{\rm max} =600$~MeV ($q_{\rm max}$ for $|\vec{q}~|$ in the $d^3q$ integration of the loop $G$ function ) as done in Refs.~\cite{daixie,liang}.
Since with these channels we only get scattering matrices in the neutral states, taking into account the $I=1$ character of all these amplitudes with involve $\pi \eta$, and the phase convention of the isospin multiples ($K^+, K^0$), ($\bar K^0, -K^-$), ($-\pi^+, \pi^0, \pi^-$), we find
\begin{equation}
t_{K^+ \bar K^0, \pi^+ \eta } = \sqrt{2} t_{K^+ K^-, \pi^0 \eta}; ~~~~
t_{\pi^+ \eta, \pi^+ \eta} = t_{\pi^0 \eta, \pi^0 \eta}.
\end{equation}

Another small technical detail is that since our amplitudes are good up to about $M_{\rm inv} = 1200$~MeV, we make a smooth extrapolation of $Gt$ at higher energies, as done in Refs.~\cite{vinicius,raquel,toledoikeno} and the results barely change for different sensible extrapolations.

\section{results}
We have seen that we have four parameters $A$, $\beta$, $B$, $\gamma$. $A \beta $ gives the strength of the hadronization of the upper vertex in external emission (Fig.~\ref{fig:3}(a) plus 3(c)). $A$ the strength for hadronization of the lower $q \bar q$ components of external emission (Fig.~\ref{fig:3}(b) plus 3(d)). $B$ the strength for the hadronization of the upper $q \bar q$ pair in internal emission (Fig.~\ref{fig:4}(a) plus 4(c)) and $B\gamma$ the strength for hadronization of the lower $q \bar q$ component in internal emission (Fig.~\ref{fig:4}(b) plus 4(d)).
One parameter, we take $A$ for this purpose, provides an arbitrary normalization and we take it $1$ or $-1$, since we only evaluate the shapes of the distributions and the ratios of the widths for $D^+ \to \pi^+ \eta \eta$  and $D^+ \to \pi^+ \pi^0 \eta$. For the second reaction, the PDG~\cite{pdg} provides the branching ratio obtained from CLEO~\cite{arun}
\begin{equation}
 {\cal B} (D^+ \to \pi^+ \pi^0 \eta) = (1.38 \pm 0.35) \times 10^{-3}.
\label{eq:BR_CLEO}
\end{equation} 

The $D^+ \to \pi^+ \eta \eta$ is measured for the first time in BESIII~\cite{besdecay}, where also the $D^+ \to \pi^+ \pi^0 \eta$ decay is studied and the following branching ratios, based on improved statistics, are reported
\begin{eqnarray}
 {\cal B} (D^+ \to  \pi^+ \eta \eta) &=& (2.96 \pm 0.24 \pm 0.10) \times 10^{-3}, \\
 {\cal B} (D^+ \to  \pi^+ \pi^0 \eta)& =& (2.23 \pm 0.15 \pm 0.10) \times 10^{-3}.
\label{eq:BR_BES}
\end{eqnarray}
We can see that the result for ${\cal B}(D^+ \to \pi^+ \pi^0 \eta)$ of CLEO (Eq.~(\ref{eq:BR_CLEO})) and BESIII (Eq.~(\ref{eq:BR_BES})) are different, and even incompatible counting errors, although they are close.

Our strategy is to find a set of three parameters $\beta$, $B$, $\gamma$ that provide a ratio of 
\begin{equation}
R = {\cal B} (D^+ \to  \pi^+ \eta \eta)/{\cal B} (D^+ \to  \pi^+ \pi^0 \eta)
= 1.33 \pm 0.16 ,
\label{eq:ratio}
\end{equation}
summing relative errors in quadrature. We shall then search for parameters that give the ratio of Eq.~(\ref{eq:ratio}) between $1-2$ if possible. First we start from $A=1$ and take values of $B$, $\beta$, $\gamma$ in the range:
\begin{equation}
 A =1;~~~ B \in [-0.2, 0.6];~~~ \beta \in [-1, 1.5];~~~ \gamma \in [0.5, 1.5].
\label{eq:para_range}
\end{equation}
The reason for the range of parameters is that we expect $B$ to be suppressed by the number of colors, and then $B\sim \frac{1}{3}$.
The relative sign between external emission and internal emission is favored to be positive in analyses of $\Lambda_c \to p \pi^+ \pi^-$ measured at BESIII~\cite{Wang:2020pem} and $B^+ \to J/\psi \omega K^+$ measured by the LHCb collaboration~\cite{Dai:2018nmw}. Yet the present process is different, and as in Ref.~\cite{enwang} we shall explore the results with opposite sign. We should also warn at this point that the $D^+ \to \pi^+ \pi^0 \eta$ process will have contribution from $\rho^+ \eta$.
This can come from the diagram of Fig~\ref{fig:1}(b) when the $\bar d u$ pair becomes a $\rho^+$ meson and $d \bar d \equiv -\frac{\pi^0}{\sqrt{2}} + \frac{\eta}{\sqrt{3}}$. The factor $\frac{1}{\sqrt{3}}$ will cause a reduction of this mechanism, but the absence of hadronization of the $\rho^+ \eta$ channel can compensate for that and we anticipate a clear contribution of the $\rho^+ \eta$ channel comparable to the case that we evaluate based on $s$-wave process that will generate the $a_0(980)$ resonance with a fairly large strength. However, there is a way to remove the $\rho^+ \eta$ contribution, imposing a cut $M_{\rm inv}(\pi^+ \pi^0) > 1$~GeV, which was already used in $D^+_s \to \pi^+ \pi^0 \eta$~\cite{besds}, which facilitates the comparison with our results. This was the case when comparing those results of Ref.~\cite{besds} with theory based upon only $s$-wave interaction in Ref.~\cite{raquel}, where a very good agreement was found. Unlike the work of Ref.~\cite{raquel}, based on only internal emission, which had no free parameters up to an arbitrary normalization, the present reactions contain both external and internal emission and we have three free parameters. 

\begin{figure}[tb!]
 \begin{minipage}[t]{0.48\hsize}
 \includegraphics[width=1.05\textwidth]{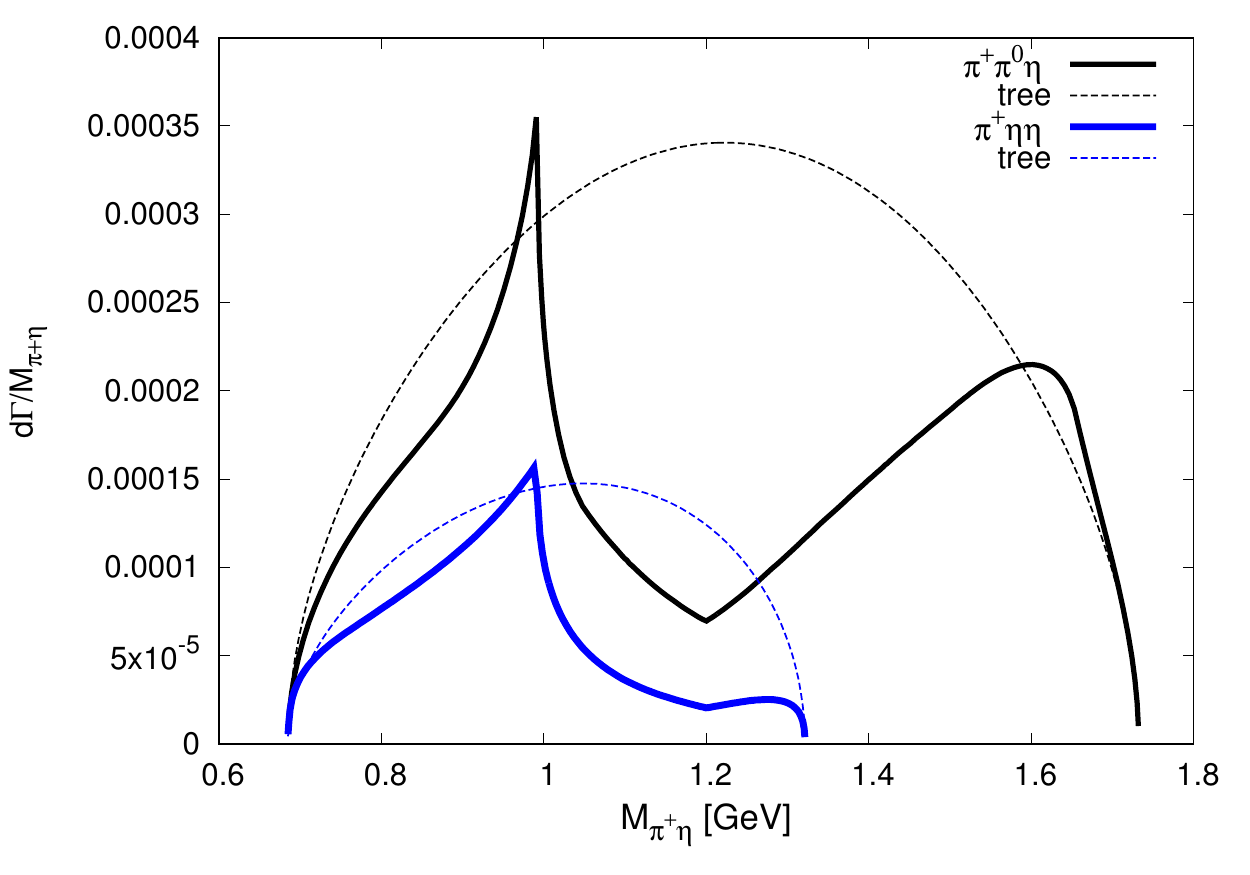}
 \vspace{-5mm}
 \caption{Solid lines: Results for ${d\Gamma}/{dM_{\rm inv} (\pi^+\eta)}$ for $D^+ \to \pi^+ \pi^0 \eta$ (upper curve) and $D^+ \to \pi^+ \eta \eta$ (lower curve). The dashed curves stand for the phase space of $D^+ \to \pi^+ \pi^0 \eta$ (upper curve) and $D^+ \to \pi^+ \eta \eta$ (lower curve). 
}
 \label{fig:new}
\end{minipage}
 \hspace{2mm}
\begin{minipage}[t]{0.48\hsize}
 \includegraphics[width=1.05\textwidth]{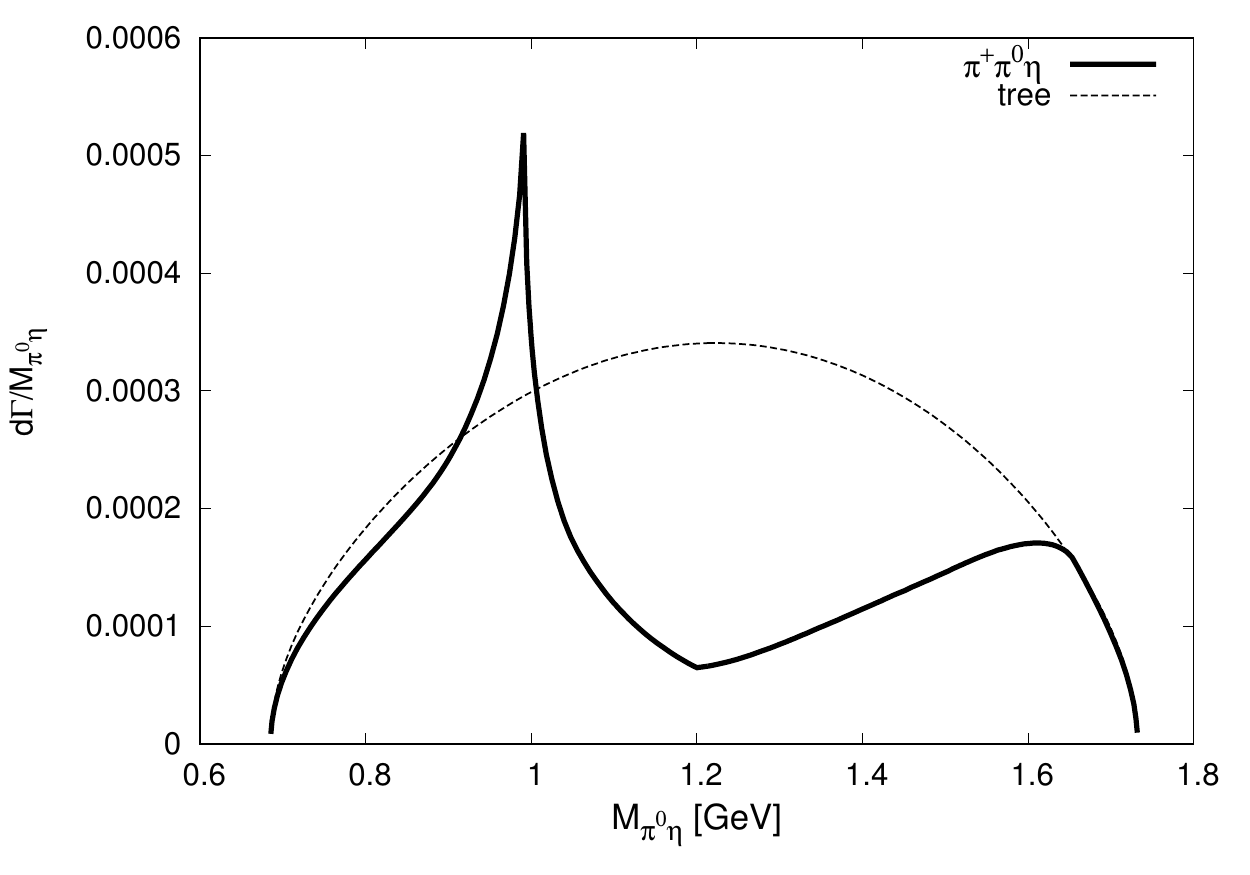}
 \vspace{-5mm}
 \caption{
Solid line: Result for ${d\Gamma}/{dM_{\rm inv} (\pi^0\eta)}$ for $D^+ \to \pi^+ \pi^0 \eta$. The dashed curve stands for the phase space of $D^+ \to \pi^+ \pi^0 \eta$.}
 \label{fig:new2}
\end{minipage}
 \end{figure}

Before we do the fit to the data, we find illustrative to see our results with a standard set of parameters: $A =1$, $\beta=1$, $B=\frac{1}{3}$, $\gamma = 1$. The results of $\frac{d\Gamma}{dM_{\rm inv}}$ for $D^+ \to \pi^+ \pi^0 \eta$ and $D^+ \to \pi^+ \eta \eta$  are shown in Fig.~\ref{fig:new}. We also plot there the shape of the phase space calculated taking only the term 1 in Eqs.~(\ref{eq:T_pipieta}), (\ref{eq:T_pietaeta}). What we see is that the difference between the shape of phase space and the results of our model is huge, indicating the important role played by the final state interaction of the meson components originated in the first step. This should be sufficient to show the value of these reactions to provide information on the meson-meson interaction.
In Fig.~\ref{fig:new2}, we also show the results with the same parameter for the $\pi^0 \eta$ distribution in the $D^+ \to \pi^+ \pi^0 \eta$ decay. 
The strength at the peak is similar but not the same, which is clear from Eq.~(\ref{eq:T_pipieta}) but 
the features are qualitatively similar. From now on we shall only plot results for the $\pi^+ \eta$ distribution.

\begin{figure}[tb!]
\begin{minipage}[t]{0.48\hsize}
 \centering
 \includegraphics[width=1.05\textwidth]{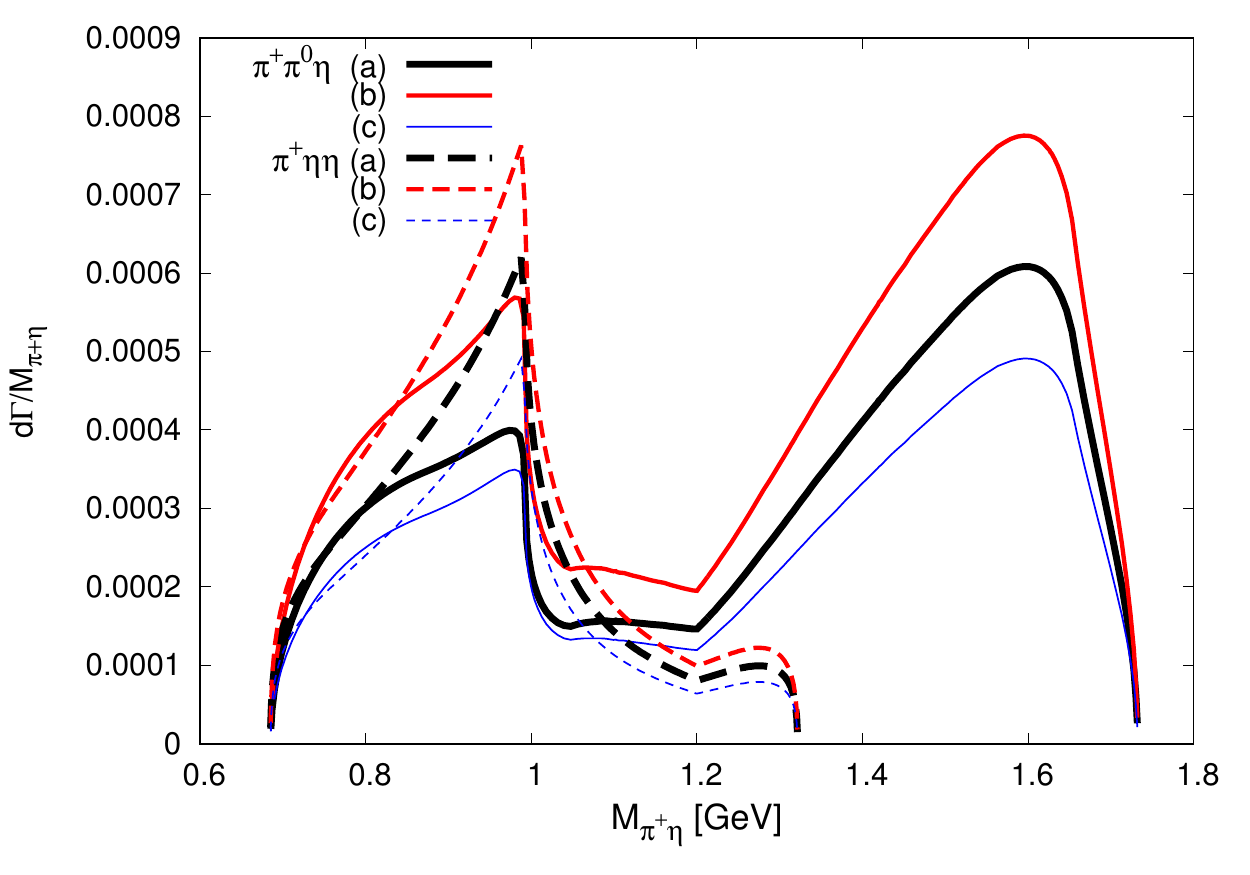}
 \vspace{-5mm}
 \caption{Differential cross sections for $D^+ \to \pi^+ \pi^0 \eta$ (solid lines) and$D^+ \to \pi^+ \eta \eta$ (dashed lines). The lables (a), (b), (c) stand for the sets of parameters of Eq.~(\ref{eq:para1}). }
 \label{fig:7}
\end{minipage}
\hspace{2mm}
 \begin{minipage}[t]{0.48\hsize}
 \centering
 \includegraphics[width=1.05\textwidth]{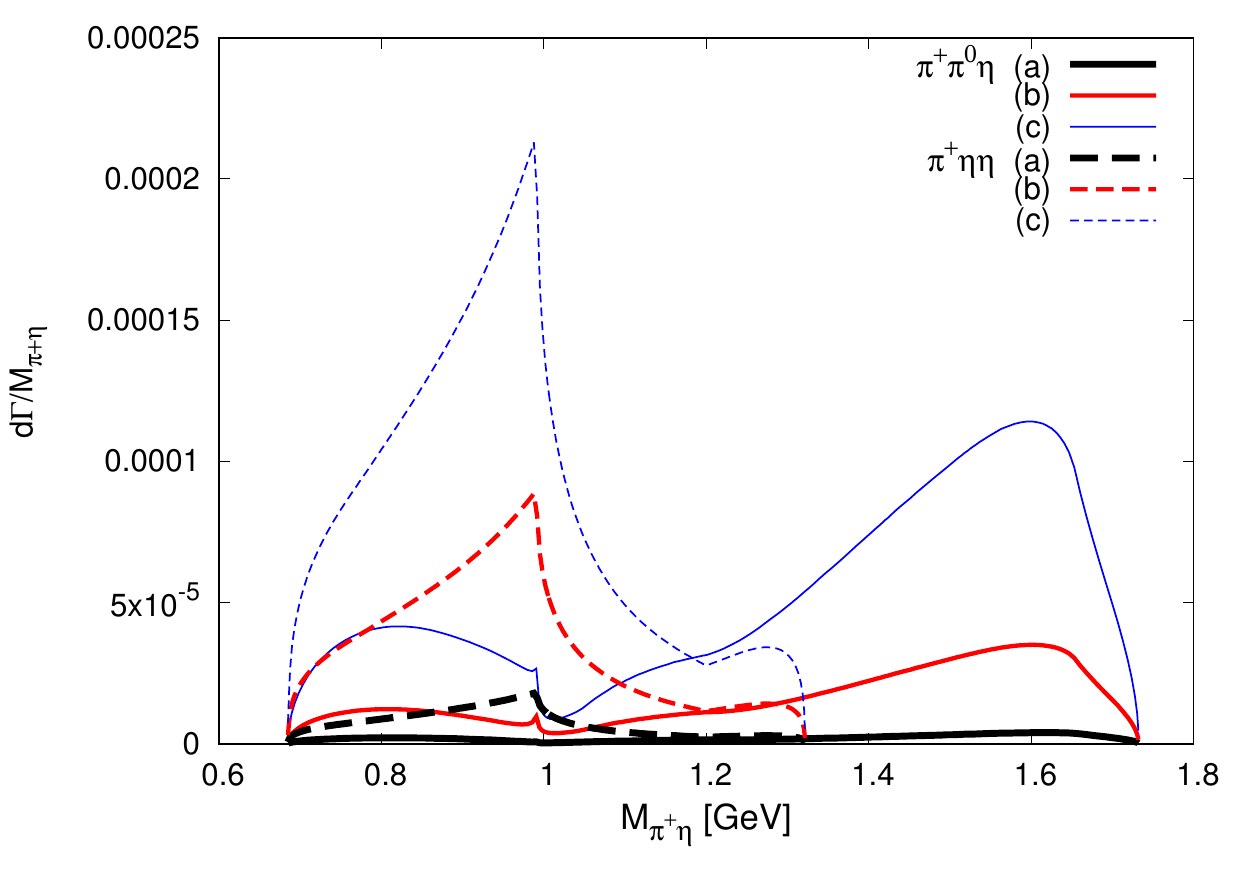}
 \vspace{-5mm}
 \caption{ Same as in Fig.~\ref{fig:7}. The labels (a), (b), (c) stand for the sets of parameters of Eq.~(\ref{eq:para2}).}
 \label{fig:8}
\end{minipage}
\end{figure}

With the caveats discussed above, we make a survey of the ratio $ R$ with different parameters at the range of  Eq.~(\ref{eq:para_range}) and we observe that it is impossible to get a ratio bigger than 0.5, about a factor three lower than the ratio of Eq.~(\ref{eq:ratio}).  One of the reasons is that the $D^+ \to \pi^+ \eta \eta$ reaction has a more reduced phase space than $D^+ \to \pi^+ \pi^0 \eta$. In Fig.~\ref{fig:7}, we show the results for $d \Gamma/ dM_{\rm inv}(\pi^+ \eta)$ with the sets of parameters,
\begin{eqnarray}
& & (a)~~\beta = 3.0,~~~ B= 0.15,~~~ \gamma=0.33,~~~ R =0.46,\nonumber\\
& & (b)~~\beta = 3.0,~~~ B= 0.55,~~~ \gamma=0.33,~~~ R =0.44,\nonumber\\
& & (c)~~\beta = 2.6,~~~ B= 0.15,~~~ \gamma=0.33,~~~ R =0.45.
\label{eq:para1}
\end{eqnarray}

In Fig.~\ref{fig:7}, we see that the $D^+ \to \pi^+ \eta \eta$ reaction has a neat $a_0(980)$ signal, with the sharp peak corresponding to the cusp-like shape of the $a_0(980)$.
The shapes for $D^+ \to \pi^+ \pi^0 \eta$ are rather different, with a broad bump at low invariant masses, caused by the tree level contribution, and much strength at high invariant masses, caused again by the tree level and the $\pi^0 \eta$ producing the $a_0(980)$. This broad contribution at higher invariant masses was also visible in the $D_s \to \pi^+ \pi^0 \eta$ experiment of Ref.~\cite{besds}. This large contribution, in a region of phase space not allowed in $D^+ \to \pi^+ \eta \eta$ must be seen as the main reason on why it is difficult to get ratio $R$ bigger than 0.5.

The important point concerning the results of Fig.~\ref{fig:7}, is to note that in all cases and for the two reactions the peak of the $a_0(980)$ is clearly seen, and in our approach we have not introduced the resonance by hand, but comes automatically from the rescattering of the particles, and not only the final particles observed, but also rescattering of $K \bar K$ pairs produced in a first step. This conclusion is the same one reached in Ref.~\cite{enwang}.

In view of the impossibility to obtain a ratio $R$ bigger than 0.5 with $A=1$, we take now $A=-1$ with the other parameters in the same range of Eq.~(\ref{eq:para_range}). In this case we can obtain bigger ratios than before. We select three parameter sets,
\begin{eqnarray}
& & (a)~~\beta = 0.72,~~~ B= 0.6,~~~ \gamma=1.21,~~~ R =2.22, \nonumber\\
& & (b)~~\beta = 1.48,~~~ B= 0.6,~~~ \gamma=1.50,~~~ R =1.35, \nonumber\\
& & (c)~~\beta = 2.24,~~~ B= 0.6,~~~ \gamma=1.50,~~~ R =1.00.
\label{eq:para2}
\end{eqnarray}

The mass distributions for these sets are shown in Fig.~\ref{fig:8}. The case (a) in Eq.~(\ref{eq:para2}) leads to a big ratio $R =2.22$ but at the expense of having very small individual rates due to large cancellations. What we observe here is that the shape of the $a_0(980)$ is much more clearly seen in the $D^+ \to \pi^+ \eta \eta$ reaction than in the $D^+ \to \pi^+ \pi^0 \eta$ one. In this latter case the $a_0(980)$ is barely seen as a small cusp at the $K \bar K$ threshold.


What we see is that the shape of the $\pi^+ \eta$ mass distributions depends strongly one the set of parameters that we take. The other reading of this finding is that the measurement of the mass distributions in both reactions, which is not done so far, will provide information on the reaction mechanisms for these two decays. The remarks obtained here should be a motivation for further measurement of the reactions, with larger statistics that allow the mass distributions to be obtained.

\begin{figure}[tb!]
 \begin{minipage}[t]{0.48\hsize}
 \centering
 \includegraphics[width=1.05\textwidth]{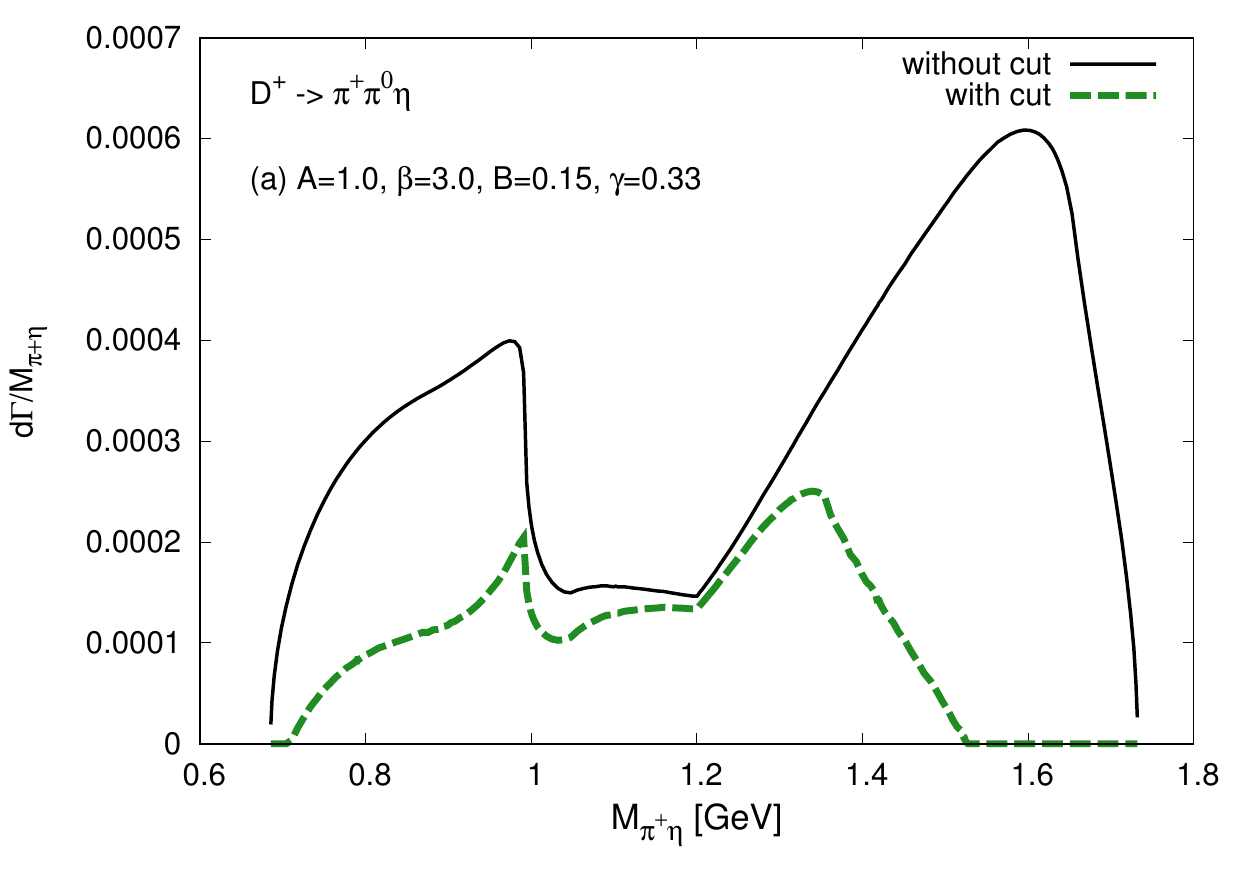}
 \vspace{-5mm}
\caption{ Results for $D^+ \to \pi^+ \pi^0 \eta$ with and without the cut $M_{\rm inv}(\pi^+ \pi^0)>1$~GeV for the set $A=1$, $\beta =3$, $B=0.15$, $\gamma=0.33$.
}
 \label{fig:9}
\end{minipage}
\hspace{2mm}
 \begin{minipage}[t]{0.48\hsize}
 \centering
 \includegraphics[width=1.05\textwidth]{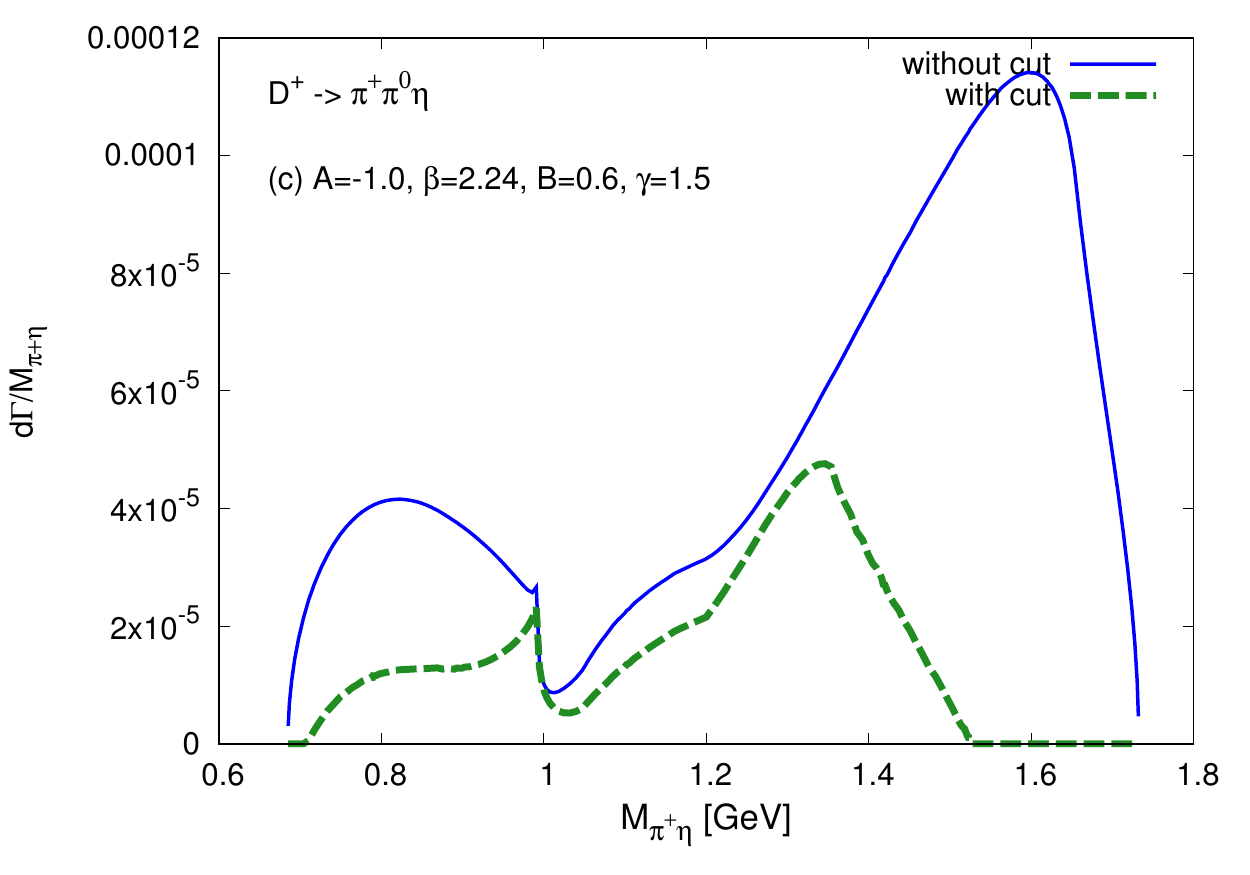}
 \vspace{-5mm}
 \caption{ Same as Fig.~\ref{fig:9} for Fig.~\ref{fig:8}, $A=-1$, $\beta =2.24$, $B=0.6$, $\gamma=1.5$.
}
 \label{fig:10}
\end{minipage}
\end{figure}

To finalize the work, we show in Figs.~\ref{fig:9} and~\ref{fig:10} the results for two selected distributions of Figs.~\ref{fig:7} and~\ref{fig:8} when we make a cut demanding that $M_{\rm inv}(\pi^+ \pi^0)> 1$~GeV. Since our variables are $M_{\rm inv}(\pi^+ \eta)$, $M_{\rm inv}(\pi^0 \eta)$, we use the relationship,
\begin{equation}
 M^2_{\rm inv}(\pi^+ \eta) + M^2_{\rm inv}(\pi^0 \eta) + M^2_{\rm inv}(\pi^+ \pi^0) =
M^2_{D^+} + m^2_{\pi^+} + m^2_{\pi^0} + m^2_\eta .
\end{equation}

In Fig.~\ref{fig:9}, we choose the results of Fig.~\ref{fig:7} for set (a) of Eq.~(\ref{eq:para1}) for $D^+ \to \pi^+ \pi^0 \eta$ and plot the corresponding results without the cut and with the cut of $M_{\rm inv}(\pi^+ \pi^0) > 1$~GeV. We can see that the effect is a reduction of the contribution at large $M_{\rm inv}(\pi^+ \eta)$, similar to what was found in Ref.~\cite{enwang}, and also at low invariant masses.

In Fig.~\ref{fig:10}, we show the same results for the set (c) of Fig.~\ref{fig:8}, the parameters of (c) of Eq.~(\ref{eq:para2}). The effect of the cut is similar to the former case, with the reduction of the strength at larger $M_{\rm inv}(\pi^+ \eta)$, and also at low invariant masses.

\section{Conclusions}

We have studied the $D^+ \to \pi^+ \eta \eta$ and $D^+ \to \pi^+ \pi^0 \eta$ reactions from a perspective where the possible mechanisms at the quark level are considered with unknown strength. Both reactions are single Cabibbo suppressed and we find that both the internal and external emission mechanisms are possible. After this, hadronization of the $q \bar q$ pairs of the different mechanisms is considered, keeping not only $\pi^+ \eta \eta$ or $\pi^+ \pi^0 \eta$ but also other intermediate states which upon rescattering can produce these final states. The pairs of mesons obtained after the hadronization are let to interact, using the chiral unitary approach to account for that interaction. The reactions selected are relevant because one can only produce the $a_0(980)$ resonance upon rescattering and the $f_0(500)$ and $f_0(980)$ do not show up. This makes the reactions useful to learn about the interaction of pseudoscalar mesons in the scalar sector with isospin 1.

In order to constrain the values of the parameters we used the ratio $R$ of the branching ratios of $D^+ \to \pi^+ \eta \eta$ and $D^+ \to \pi^+ \pi^0 \eta$. This indeed puts much constraints on the parameters, but we still had freedom to obtain values of $R$ bigger from 1 for different sets, leading to different shapes of the $\pi \eta$ mass distributions. We observed that, in all cases the $a_0(980)$ signal was visible in the $\pi^+ \eta$ invariant mass distributions, but was more neat in the $D^+ \to \pi^+ \eta \eta$ reaction. We conclude that, while we clearly expect to see the $a_0(980)$ signal in the $\pi \eta$ mass distributions, there are still uncertainties in the theory concerning the actual shape tied to the details of the reaction mechanism. In this sense, when the actual mass distributions are measured, information will be available that allows us to pin down the reaction mechanisms and come out with more assertive conclusions on the role played by the $a_0(980)$ resonance in these reactions.

\section*{ACKNOWLEDGEMENT}
We wish to thank En Wang for useful discussion.
The work of N. I. was partly supported by JSPS Overseas Research Fellowships and JSPS KAKENHI Grant Number JP19K14709. This work is partly supported by the Spanish Ministerio de Economia y Competitividad and European FEDER funds under
Contracts No. FIS2017-84038-C2-1-P B and No. FIS2017-84038-C2-2-P B. This project has received funding from
the European Unions Horizon 2020 research and innovation programe under grant agreement No 824093 for the
**STRONG-2020 project.

\end{document}